\documentclass[12pt]{article}
\begin{document}
\begin{center}
 A LAGRANGIAN FOR DSR PARTICLE AND THE ROLE OF NONCOMMUTATIVITY  \vskip 2cm

Subir Ghosh\\
Physics and Applied Mathematics Unit,\\
Indian Statistical Institute,\\
203 B. T. Road, Calcutta 700108, India.
\end{center}
\vskip 3cm
{\bf Abstract:}\\
In this paper we have constructed a coordinate space (or
geometric) Lagrangian  for a point particle that satisfies the exact
Doubly Special Relativity (DSR) dispersion relation in the
Magueijo-Smolin framework. Next we demonstrate how a
Non-Commutative phase space is needed to maintain Lorentz
invariance for the DSR dispersion relation. Lastly we address the very important issue of velocity of this DSR particle. Exploiting the above Non-commutative phase space algebra in a Hamiltonian framework, we show that the speed of massless particles is $c$ and for massive particles the speed saturates at $c$ when  the particle energy reaches the maximum value $\kappa $, the Planck mass.

\newpage
Motivated by ideas from quantum  gravity \cite{planck} an
extension of Special Theory of Relativity, known as  Doubly (or
Deformed) Special Relativity (DSR) \cite{am}, has been proposed. Indeed, it should be emphasized that, (as in Special Theory of Relativity), DSR is also based on the sacred (Einsteinian)  relativity principle of  inertial observers, that automatically removes the idea of a preferred reference frame (see for example \cite{gr}). However, unlike Special Theory of Relativity, that has a single observer-independent scale - the velocity of light $c$ - in DSR there are {\it{two}} observer-independent scales: a length scale, 
($\sim$ Planck length?) and  the velocity of light $c$.  This
construction generalizes the conventional energy-momentum
dispersion relation to $p^2=m^2 +F(\kappa , ..)$ where $m$ is the
rest mass and the extension function $F$ depends on a new
parameter $\kappa $ (besides the existing variables and
parameter). $\kappa $ is related  (may be) to the Planck mass.
However, in the limit $\kappa \rightarrow \infty$ one recovers the
relation $p^2=m^2$.

This additional mass scale $\kappa $ plays crucial roles in two
very different contexts: The new dispersion relation is useful in
explaining observations of ultra-high energy cosmic ray particles and photons,
that violate the GZK bound \cite{am1}. Note that one can "solve" these Threshold Anomaly problems by introducing explicitly Lorentz symmetry violating schemes but this is possible only at the cost abundoning the Relativity Principle. At the same time, existence
of a length scale is directly linked to the breakdown of
spacetime continuum and the emergence of a Non-Commutative (NC)
spacetime (below {\it{e.g.}} Planck length) \cite{dfr,sn,rev}. Once again, this is in accord with the Relativity Principle since all inertial observers should agree to the (energy) scale that signals the advent of new physics, (maybe) in the form of an underlying NC spacetime structure and its related consequences. In
this paper we will focus on the second aspect, after constructing
a dynamical model for a DSR particle.

The theoretical development in this area so far has been mainly
kinematical in the sense that various forms of the generalized
dispersion relation ({\it{i.e.}} forms of $F$) have been suggested that
are consistent with $\kappa $-dependent extentions of Poincare algebras
\cite{dsr2,mag,kow}. However, a satisfactory geometrical picture
of the model, in terms of a {\it{coordinate space Lagrangian}},  so far
has not appeared.  In the present work we have provided such a
Lagrangian that can describe a particular form of DSR dispersion
relation, known as the Magueijo-Smolin (MS) relation \cite{mag},
\begin{equation}
p^2=m^2[1-\frac{(\eta ^\mu p_{\mu})}{\kappa} ]^2,  \label{m}
\end{equation}
where  $\eta^0=1,\vec \eta =0$. In (\ref{m}), $p_0=\kappa $
provides the particle energy upper bound, which can be identified
with the Planck mass. A first step in model building was taken in
\cite{sg} where the system described the MS particle only for
$m=\kappa $. The  model presented here is valid for the exact MS relation
(\ref{m}).

The other important issue is the connection between this particle
model with a Non-Commutative (NC) spacetime (or more generally
phase space) \cite{rev,dsr2,mag,kow}. Exploiting the notion of
duality in the context of Quantum Group ideas, it has been
demonstrated \cite{dsr2,mag,kow} that each DSR relation is
{\it{uniquely}} associated with a particular form of NC phase
space. More precisely, a DSR relation is  Casimir of a particular
$\kappa $-deformed Poincare algebra and the latter is connected to
an NC phase space in a unique way. In particular, the MS relation
is related to a specific representation of $\kappa $-Minkowski NC
phase space \cite{dsr2,mag,kow}.

From a different perspective, one can directly obtain the phase
space algebra of a point particle model simply by studying its
symplectic structure (in a first order phase space Lagrangian
formulation \cite{fj}) or by performing a constraint analysis in a
Hamiltonian framework \cite{dirac}. The most popular example in
this connection is the canonical NC Moyal plane that one gets in
studying the planar motion of a charge in a large perpendicular
magnetic field \cite{rev}. Generation of NC phase space with Lie
algebraic forms of noncommutativity have appeared in \cite{g}.
The motivation of our work is also to see in an explicit way how
the connection between a DSR relation and a specific NC phase
space is uniquely established, in a dynamical framework, as an alternative to the
Quantum Group duality approach \cite{dsr2,mag,kow}. The
conclusion drawn from our present analysis of an explicit model is
quite interesting. We find that, for a modified dispersion
relation (such as DSR mass-energy law) appearance of a modified phase
space algebra (or an NC phase space) is {\it{necessary}} but the
association between a DSR relation and an NC phase space is
{\it{not}} unique.  The first assertion is originated from a
subtle interplay between Lorentz invariance and the DSR dispersion
relation in question. The non-uniqueness in the choice of NC phase space is due to the
gauge choice and relative strengths of the parameters. In the present work we demonstrate
that the
MS law is consistent with an NC phase space algebra that is more
general than the $\kappa $-Minkowski. This is a new and mixed form
of NC phase space algebra that interpolates between {\it{two}} Lie
algebraic structures: Snyder  \cite{sn} and $\kappa $-Minkowski (in MS base) \cite{dsr2,kow}.

The important question of three velocity  of the particle appears is answered very clearly in our dynamical framework. Our results show that the massless particles move with $c$ and the maximum speed  of massive particles is also $c$, when their energy reaches the upper bound $\kappa $ and there are subtle $\kappa$-efects for the general case. These conclusions agree with \cite{das}.

Let us start with the Lagrangian of our proposed model of an MS
particle,
$$
L=\frac{m\kappa}{\sqrt{\kappa^{2}-m^{2}}}[g_{\mu\nu}\dot
x^{\mu}\dot x^{\nu}+\frac{m^2}{\kappa^{2}-m^{2}}(g_{\mu\nu}\dot
x^{\mu}\eta^{\nu})^2]^{\frac{1}{2}}-\frac{m^2\kappa}{\kappa^{2}-m^{2}}g_{\mu\nu}\dot
x^{\mu}\eta^{\nu}$$
\begin{equation}
 \equiv
\frac{m\kappa}{\sqrt{\kappa^{2}-m^{2}}}\Lambda -
\frac{m^2\kappa}{\kappa^{2}-m^{2}}(\dot x\eta ).
 \label{01}
\end{equation}
Here $g_{\mu\nu}$ represents  the flat Minkowski metric
$g_{00}=-g_{ii}=1$. We have adopted the shorthand notation
$(AB)=g_{\mu\nu}A^{\mu}B^{\nu}$ and $c=1$..

First we derive the DSR dispersion relation.  The conjugate
momentum is
\begin{equation}
p_{\mu}\equiv \frac{\partial L}{\partial \dot
x^\mu}=\frac{m\kappa}{\sqrt{\kappa^{2}-m^{2}}}\frac{(\dot x_\mu
+\frac{m^2}{\kappa ^2-m^2}(\dot
x\eta)\eta_{\mu})}{\Lambda}-\frac{m^2\kappa}{\kappa^{2}-m^{2}}\eta^{\nu}.
\label{m4}
\end{equation}
It is straightforward to check that (\ref{m4}) satisfies the MS
dispersion law (\ref{m}). The structure of a point particle model
of the kind (\ref{01}) is new and is one of our main results. We note that the last term (although being a total derivative) and the specific overall scale factor in (\ref{01}) is required to yield the MS relation (\ref{m}).

Let us now discuss why the  NC phase space is necessary. To begin
with, one can construct the above Lagrangian from the first order
form,
\begin{equation}
L=(\dot x p)-\frac{\lambda}{2}[p^2-m^2(1-\frac{(\eta p)}{\kappa}
)^2],
 \label{02}
\end{equation}
by eliminating $\lambda $ and $p_\mu $. In (\ref{02}) $\lambda $
plays the role of a multiplier that enforces the MS  mass-shell
condition. The symplectic structure in (\ref{02}) clearly suggests
a canonical phase space with the only non-trivial Poisson bracket
$\{x_{\mu},p_{\nu}\}=-g_{\mu\nu}$. But notice that the MS law is
{\it{not}} compatible with  Lorentz invariance if one employs  a
canonical phase space. Quite obviously the Lorentz generator
$J_{\mu\nu}=x_\mu p_\nu - x_\nu p_\mu $ transforms $x_\sigma $ and
$p_\sigma $ properly,
\begin{equation}
\{J_{\mu\nu},x_\sigma \}=g_{\nu \sigma }x_\mu -g_{\mu \sigma
}x_\nu ;~~\{J_{\mu\nu},p_\sigma \}=g_{\nu \sigma }p_\mu -g_{\mu
\sigma }p_\nu ,
 \label{lor}
\end{equation}
but it fails to keep the MS dispersion law invariant,
\begin{equation}
\{J_{\mu\nu}, (p^2-m^2(1-\frac{(\eta p)}{\kappa})^2)
\}=-\frac{2}{\kappa }(1-\frac{(\eta p)}{\kappa})(\eta _\mu p_\nu
-\eta _\nu p_\mu ).
 \label{lor1}
\end{equation}
The remedy is to introduce a modified or NC phase space algebra
that is consistent with the present Lagrangian structure
(\ref{01}) and keeps the MS relation invariant.

This is possible thanks to the $\tau$-reparameterization
invariance of  the Lagrangian (\ref{01}), which is evident from
the vanishing Hamiltonian,
\begin{equation}
 H=(p\dot x)-L=0.
\label{hh}
\end{equation}
 This local gauge invariance allows us to choose appropriate gauge
fixing conditions such that specific forms of NC phase space
structures are induced via Dirac Brackets \cite{dirac}. In the
terminology of Dirac, the non-commuting constraints are termed as
Second Class Constraints (SCC) and the commuting constraints, that
induce local gauge invariance, are First Class (FCC). In the
presence of  SCCs $(\psi_{1},\psi_{2})$ that do not
commute, $\{\psi_1 ,\psi _2 \}\ne 0$, the  Dirac Brackets are
defined in the following way,
\begin{equation}
\{A,B\}^*=\{A,B\}-\{A,\psi _i\}\{\psi _i,\psi _j\}^{-1}\{\psi
_j,B\}, \label{n4}
\end{equation}
where $\{\psi _i,\psi _j\}$ refers to the constraint matrix. From
now on we will always use Dirac brackets and refer them simply as
$\{A,B\}$.  In the present instance, the MS mass-shell condition
(\ref{m}) is the only FCC present and there are no SCC. We choose
the  gauge,
\begin{equation}
 \psi _1 \equiv (xp)=0,
\label{di}
\end{equation}
that has been considered  before \cite{g,gauge} in similar
circumstances. Together with the mass-shell condition (\ref{m}), 
$\psi _2 \equiv p^2-m^2(1-(\eta p)/\kappa)^2= 0$ they constitute
an SCC \cite{dirac} pair with the only non-vanishing constraint
matrix element $\{\psi_1 ,\psi _2 \}=-m^2(1-(\eta p)/\kappa )$. Hence
the Dirac brackets follow:
$$
\{x_\mu ,x_\nu \}=\frac{1}{\kappa}(x_\mu \eta_{\nu}-x_\nu
\eta_{\mu })+\frac{1}{m^2(1-(\eta p)/\kappa)}(x_\mu p_{\nu}-x_\nu
p_{\mu }),$$
\begin{equation}
\{x_{\mu},p_{\nu}\}=-g_{\mu\nu}+\frac{1}{\kappa}\eta_{\mu}p_{\nu}+\frac{p_{\mu}p_{\nu}}{m^2(1-(\eta
p)/\kappa)},~~\{p_{\mu},p_{\nu}\}=0.
 \label{03}
\end{equation}
Performing an (invertible) transformation on the variables,
\begin{equation}
\tilde x_\mu =x_\mu -\frac{1}{\kappa }(x\eta )p_\mu ,
 \label{04}
\end{equation}
we find  an interesting form of algebra that interpolates
between  Snyder \cite{sn} and $\kappa $-Minkowski
\cite{am,dsr2,kow}:
$$
\{\tilde x_\mu , \tilde x_\nu \}=\frac{1}{\kappa }(\tilde x_\mu
\eta _\nu -\tilde x_\nu \eta _\mu )+\frac{\kappa ^2 -m^2}{\kappa
^2 m^2}(\tilde x_\mu p_\nu -\tilde x_\nu p _\mu ),$$
\begin{equation}
\{\tilde x_\mu , p_\nu \}= -g_{\mu\nu}+\frac{1}{\kappa }(p_\mu
\eta _\nu +p_\nu \eta _\mu )+\frac{\kappa ^2 -m^2}{\kappa ^2 m^2}p
_\mu p_\nu ~,~\{p_\mu , p_\nu \}=0. \label{05}
\end{equation}
In absence of the $1/\kappa $-term or the $(\kappa ^2
-m^2)/(\kappa ^2 m^2)$-term, one obtains the Snyder \cite{sn} or
the $\kappa $-Minkowski algebra \cite{dsr2,kow}, respectively.

We now show that the novel phase space algebra (\ref{05}) is indeed
consistent with Lorentz invariance. With
$J_{\mu\nu}=\tilde x_\mu p_\nu - \tilde x_\nu p_\mu $ and using (\ref{05}),
one can easily compute,
\begin{equation}
\{J_{\mu\nu},(p^2-m^2(1-\frac{(\eta p)}{\kappa })^2)\}=2
(p^2-m^2(1-\frac{(\eta p)}{\kappa })^2) \approx \psi _2(\eta _\mu p_\nu -\eta _\nu
p_\mu ), \label{50}
\end{equation}
so that the MS relation is Lorentz invariant on-shell. Next, using (\ref{05}),  we
check that the Lorentz algebra is intact,
\begin{equation}
\{J^{\mu\nu },J^{\alpha\beta }\}=g^{\mu\beta }J^{\nu\alpha
}+g^{\mu\alpha }J^{\beta \nu}+g^{\nu\beta }J^{\alpha\mu
}+g^{\nu\alpha }J^{\mu\beta }. \label{51}
\end{equation}
This, in itself, is expected since individually both Snyder
\cite{sn} and $\kappa $-Minkowski \cite{dsr2} algebras does not
modify the Lorentz sector, but all the same it is reassuring to
note that the mixed form (\ref{05}) also has this property. However,
Lorentz transformations of $x_\mu$ and $p_\mu $ are indeed
affected,
\begin{equation}
\{J^{\mu\nu},\tilde x^\rho \}=g^{\nu\rho}\tilde x^\mu-g^{\mu\rho}\tilde x^\nu
-\frac{1}{\kappa } (\eta^\mu J^{\rho\nu}-\eta^\nu J^{\rho\mu })~;~
\{J^{\mu\nu},p^\rho \}=g^{\nu\rho}p^\mu -g^{\mu\rho}p^\nu
-\frac{1}{\kappa }(\eta^\nu p^\mu-\eta^\mu p^\nu)p^\rho .
\label{52}
\end{equation}
Notice that only the $\kappa $-Minkowski part of the algebra
(\ref{05}) is responsible for the above modified forms and also that
the extra terms appear only for $J^{0i}$ and not for $J^{ij}$ so
that only boost transformations are changed. This concludes our discussion on the construction of MS particle Lagrangian and its associated NC phase space.

We now address the very important issue of speed of the $\kappa $-particle \cite{das}. We stress that, since we have an explicit Lagrangian construction, the definition of velocity is very natural and unambiguous in this scheme. We extract the Hamiltonian $ p_0$ from the MS mass shell condition (\ref{m}),
\begin{equation}
p_0=\frac{\kappa}{(\kappa ^2-m^2)}(-m^2+(\kappa ^2m^2+(\kappa ^2-m^2)\vec p^2)^{\frac{1}{2}})
\label{h}
\end{equation}
with $p_0\sim \sqrt{m^2+\vec p^2}$ as $\kappa \rightarrow \infty $. Next, exploiting the NC algebra (\ref{05}), we derive the particle dynamics:
\begin{equation}
\dot {\tilde x_0}\equiv \{\tilde x_0,p_0\}=\frac{\vec p^2}{m^2}~;~~\dot {\tilde x_i}\equiv \{\tilde x_i,p_0\}=\sqrt{(1+\frac{(\kappa^2-m^2)}{\kappa ^2m^2}\vec p^2)}\frac{p_i}{m}.
\label{h1}
\end{equation}
As a consistency check, note that (\ref{h1}) can be directly read off from the $\{\tilde x_\mu,p_\nu \}$ bracket given in (\ref{05}). The natural definition for three velocity  \cite{das}, $v_i\equiv \dot {\tilde x_i}/\dot {\tilde x_0}$, is not naively applicable in the present case as it does not lead to normal particle  velocity in the $\kappa \rightarrow \infty $ limit. However, this is not surprising since we have used a non-standard gauge choice (\ref{di}) and further redefinitions (\ref{04}). Let us insist that all the physical quantities in the limit $\kappa ,\kappa m \rightarrow \infty $  should reduce to normal particle properties since then the algebra (\ref{05}) becomes completely canonical. Keeping this in mind, we define a new variable,  
\begin{equation}
X\equiv (\frac{(\kappa^2-m^2)}{\kappa ^2}+\frac{m^2}{\vec p^2})\tilde x_0, 
\label{h2}
\end{equation}
and hence obtain,
\begin{equation}
v_i\equiv \dot {\tilde x_i}/\dot X=p_i/\sqrt{(m^2+\frac{(\kappa^2-m^2)}{\kappa ^2}\vec p^2}~,~~
\vec v^2=\vec p^2/(m^2+\frac{(\kappa ^2-m^2)}{\kappa ^2}\vec p^2 ).
\label{h3}
\end{equation}
First of all, we justify our choice of $X$ by noting that 
\begin{equation}
\{X,p_0\}=1+\frac{(\kappa ^2-m^2)}{\kappa ^2m^2}\vec p^2\approx 1+O(\frac{1}{\kappa ^2m^2}),
\label{t}
\end{equation}
so that in the canonical limit $X$ behaves as a conjugate variable to $p_0$, the Hamiltonian, as it aught to. The velocity in (\ref{h3}) has the correct $\kappa  \rightarrow \infty $ canonical limit. Moreover  $m^2=0\Rightarrow \vec v^2=1$ showing that massless DSR particles move with $c$ irrespective of their energy. On the other hand, for massive particles, the MS relation (\ref{m}) saturates at $p_0=\kappa $ for which $\vec p^2=\kappa ^2 $. Putting this back in (\ref{h3}) we find once again $\mid v\mid =1$. Lastly, $m=\kappa \Rightarrow \mid v\mid =\mid p\mid /m $ so that Planck mass particles appear to be non-relativistic, which agrees with their dispersion relation (\ref{m}). All these conclusions are in accord with \cite{das}. In the two figures (i) and (ii) for $m=1,\kappa =1.5$ and $m=1,\kappa =3$ respectively, we plot $\vec p^2 \equiv A(x),~\vec v^2 \equiv C(x)$ against energy $p_0\equiv x$ for the MS particle and compare them with the normal particle $\vec p^2\equiv B(x),~ \vec v^2 \equiv D(x)$. The MS energy upper bound $p_0\equiv x=\kappa $ is used in the graphs. They  indicate that MS particles can survive for smaller energy than normal particle (for the same mass) and is always faster than normal particle  of same energy. However, the velocity of massive MS particle is also bounded by $c$ that occures at $p_0=\kappa $. Figure (ii) shows that, for larger $\kappa$, the MS particle tends towards its normal cousin very quickly. We emphasize that although we have worked in a particular gauge,  the above limiting results are general since they involve only $\kappa $-relativistic invariants.

It is interesting to consider generalization of the invariant
"length" $l^2$ in our geometry,
\begin{equation}
\{J^{\mu\nu},l^2\}= \{J^{\mu\nu},(\tilde x^2(1-\frac{(\eta
p)}{\kappa })^2)-\frac{(\tilde xp)^2}{m^2} )\}= \frac{2(\tilde
xp)}{m^2\kappa }(p^2-m^2(1-\frac{(\eta p)}{\kappa })^2)(\tilde
x_\mu \eta _\nu -\tilde x_\nu \eta _\mu ). \label{53}
\end{equation}
We find that in this type of phase space geometry the notion of an
absolute coordinate space length is replaced by a combination of
both $\tilde x_\mu $ and $p_\mu$ that is invariant only on-shell (for MS law)
and for $\kappa \rightarrow \infty $ one recovers the "length" for
Snyder geometry. For Snyder algebra, this is also consistent with
the interpretation,
\begin{equation}
\{ x^S_\mu , p_\nu \}= -g_{\mu\nu}+\frac{1}{ m^2}p _\mu p_\nu
\equiv -G^S_{\mu\nu }~,~~ (l^2)^s=G^S_{\mu\nu }(x^S)^\mu (x^S)^\nu
, \label{54}
\end{equation}
where, as noted before, the Snyder algebra (and metric) is
obtained from (\ref{05}) in the limit $\kappa \rightarrow \infty
$. However this interpretation does not work if one includes the
$\kappa $-Minkowski component of the algebra.

We conclude by noting that more dramatic changes in our perception are awaiting us, as and when we are able to construct a Quantum Field Theory with the underlying $\kappa $-Minkowski Non-Commutative spacetime structure and with fundamental excitations obeying  DSR kinematics. To that end, it is essential that one has a clear understanding of the physics involved in the classical and quantum mechanical scenario. We hope that the present work is a first step in this direction.

\vskip .2cm
{\it{Acknowledgement}}: It is a pleasure to thank Debajyoti
Choudhury for discussions and Theory Group, I.C.T.P., where the
present idea took shape during our visit. Also I thank Etera Livine for comments.

\end{document}